\documentclass[11pt, a4paper]{article}
\usepackage{graphicx}
\usepackage{indentfirst,csquotes}

\usepackage{amssymb,amsthm,amsmath}
\usepackage{authblk}
\usepackage{booktabs}
\usepackage{xcolor,paralist,hyperref,fancyhdr,etoolbox}
\usepackage[numbers,sort&compress]{natbib}

\hypersetup{ colorlinks=true, breaklinks=true, linkcolor=black, citecolor=blue, filecolor=black, urlcolor=black }

\usepackage{lipsum}
\usepackage[top=2.5cm, bottom=2.5cm, left=2.5cm, right=2.5cm]{geometry}
\begin{document}
\title{Direction-Dependent Quantum Transport Properties of MoS$_2$ Integrated into Biphenylene Configuration}

\author{Gozde Özbal Sargın \\ 
\small Department of Fundamental Sciences, Air NCO Vocational HE School, Turkish National Defence University, İzmir, 35415 Türkiye}

\date{\today}
\date{\today}

\maketitle
\pagestyle{plain}

\let\thefootnote\relax
\footnotetext{gozdeozbal.sargin@msu.edu.tr} 

\begin{abstract}
Motivated by the experimental realization of the two-dimensional (2D) carbon biphenylene network (BPN), the theoretical extension of the BPN topology to various groups of elements was successfully implemented. In this work, we conducted first principles and quantum transport calculations to reveal the electronic, thermal, thermoelectric performance and current-voltage (\textit{I-V}) behavior of the pristine MoS$_2$-BPN by using density functional theory (DFT) combined with Non-Equilibrium Green's Function (NEGF) formalism. At room temperature, the phonon thermal conductance is remarkably low along both the armchair and zigzag orientations, with values of 0.28 nW/K and 0.23 nW/K, respectively. The directional dependence of the electronic and thermal transport properties is clearly reflected in the thermoelectric figure of merit ($zT$) values, which reach first peaks at 0.27 and 0.19 along the armchair and zigzag directions, respectively. The current-voltage ($I-V$) characteristics demonstrate an exceptionally strong transport anisotropy, characterized by a substantial current ratio of $I_{\text{armchair}}/I_{\text{zigzag}} \approx 7 \times 10^4$.
Furthermore, while the current along the armchair direction increases steadily with the applied bias, transport along the zigzag direction is characterized by a pronounced intrinsic negative differential conductance (NDC). This contrast highlights fundamentally distinct charge transport mechanisms along the two orthogonal axes. Consequently, different directions of this BPN phase of MoS$_2$ can be tailored for distinct nanoelectronic applications, where the armchair and zigzag axes serve as a reliable current switch and an active NDC device, respectively.
\end{abstract} 
\newpage
\section{Introduction}
Since its discovery, graphene has been both a pioneer of two-dimensional (2D) materials due to its extraordinary electronic, mechanical, and thermal properties, and an ideal benchmark for comparing investigated properties of other two-dimensional (2D) materials~\cite{doi:10.1126/science.1102896,Geim2007,RevModPhys.81.109,doi:10.1126/science.1157996,doi:10.1021/nl0731872,Balandin2011}. Over the past 20 years, within the ever-expanding pool of 2D systems, graphene-related materials such as, penta-graphene, T-graphene, Net-graphene, $\Psi$-graphene, graphenylene, monolayer amorphous carbon (MAC), graphyne-based allotropes and carbon biphenylene network (C-BPN) have held a privileged position due to both their immense diversity and unparalleled performance~\cite{doi:10.1073/pnas.1416591112,PhysRevLett.108.225505,doi:10.1021/acs.jpclett.7b01364,https://doi.org/10.1002/pssb.201046583,Toh2020,Liu2017,C2TC00006G,doi:10.1126/science.abg4509}. Among them, C-BPN emerges as a superior candidate in lithium-ion batteries, sensing applications, nano-electronic devices, thermoelectric devices due to its high adsorption capability, strain tunable band gap opening and inherently anisotropic charge and thermal transport properties~\cite{PASANAJE202583,LIMA2024112673,doi:10.1021/acs.langmuir.4c00035,XU2026166402,doi:10.1021/jacs.2c02178,PhysRevMaterials.9.024003}. 
Beyond the pristine monolayer C-BPN, its heterostructures and alloy compositions, notably BN biphenylene, have emerged as highly efficient and structurally robust platforms for advanced reversible hydrogen storage~\cite{TAO2026152948,WANG2026121171}.

It has been shown that the structural transition from the hexagonal honeycomb lattice to the carbon biphenylene in square-hexagon-octagon orientation effectively breaks the intrinsic three-fold symmetry of graphene. This reconstruction of the atomic arrangement drives the system from a semimetallic behavior toward a truly metallic character. Following this lead, investigation of diverse compounds within the biphenylene porous framework has been a focus of interest. Demirci \textit{et al.} explore the stability and structural properties of group-IV elements and IV-IV, III-V, and II-VI compounds in biphenylene network~\cite{PhysRevB.105.035408}. Silicon BPN holding a metallic character is found suitable for thermoelectrics and nanoelectronics stemming from its low lattice thermal conductivity, anisotropic electronic structure and strain-tunable structure~\cite{doi:10.1021/acsaelm.2c00459,ALIDOOSTI2025100610}. Group III- nitride BPNs exhibit robust mechanical stability, and GaN and AlN-BPN are positioned as high-potential candidates for optoelectronic applications due to their electronic profiles\cite{D3CP00776F,LOPESLIMA2023107183,doi:10.1021/acsomega.4c03511}. In addition, tuning the band gap of BN-BPN is achieved through doping with various elements, and metal-decorated BN-BPNs hold a high potential for hydrogen storage~\cite{D3CP00776F,MA2025115889}. More recently, DFT and \textit{ab inito} molecular dynamics (AIMD) computations reveal the thermal and mechanical stability of GaP- and InP-BPNs from group III-V~\cite{LIU2026116374}. In parallel to these advancements, investigations into the biphenylene-like BC$_3$ reveal its high optical absorption capacity and striking anisotropic carrier mobilities, which make it a potential candidate for use in photovoltaics and nanoelectronics~\cite{D5NR04789G}.  Clearly, while the biphenylene geometry ensures thermal and mechanical stability across various elements and groups, it simultaneously gives rise to a wide spectrum of distinct physical properties. Moreover, precise control over their electronic characteristics can be achieved through doping and strain engineering. In addition to the groups discussed earlier, Transition Metal Dichalcogenides (TMDs) such as MoS$_2$ and MoSe$_2$ have also been integrated into the biphenylene network, and their stability and fundamental electronic properties have been comprehensively elucidated~\cite{doi:10.1021/acs.jpcc.3c00388,BIRCAN2025417956}. 

Herein, we conduct a comprehensive analysis of the ballistic transport properties of pristine MoS$_2$-BPN to evaluate its potential for TE devices and nanoelectronic applications. Ballistic thermal transport calculations reveal that phonon thermal conductance at 300~K is notably suppressed in both directions, yielding values of 0.28 nW/K for the armchair and 0.23 nW/K for the zigzag configurations. Analogues to phonon thermal conductance, directional anisotropy in electronic TE coefficients results in different $p$- and $n-type$ $zT$ values of 0.27/0.26 and 0.19/0.20 in armchair and zigzag directions, respectively. Alongside its thermoelectric performance, the current-voltage characteristics of MoS$2$-BPN reveal exceptional potential for nanoscale device integration. The material displays remarkably high current magnitudes along the armchair axis and a considerable transport anisotropy ($I_{\text{armchair}}/I_{\text{zigzag}} \approx 7 \times 10^4$), suggesting a natural current switching mechanism. In contrast, the zigzag direction demonstrates a strong intrinsic negative differential conductance. This intrinsic dual-functionality positions MoS$_2$-BPN as a highly promising, manufacturing-friendly candidate for next-generation logic and oscillator devices.

\section{Method}
\label{Method_sec}
In this study, transport simulations equipped with density functional theory calculations are achieved through QuantumATK software~\cite{Smidstrup_2020}. Prior to the electronic and thermal transport calculations, structural optimization is performed by utilizing LCAO calculator, which applies numerical LCAO Linear Combination of Atomic Orbitals) basis sets. The initial lattice parameters and atomic positions were subjected to a full equilibrium search within the Generalized Gradient Approximation (GGA-PBE). To represent core-electron interactions, norm-conserving pseudopotentials from the PseudoDojo database were employed~\cite{VANSETTEN201839}. The convergence criteria for this structural equilibration were set to a maximum atomic force of 0.001 eV/Å and a pressure threshold of 0.05 GPa. A dense Monkhorst-Pack k-grid of 20 × 12 × 1 was employed to ensure high numerical accuracy during the optimization, while van der Waals forces were accounted for via the Grimme D3 method~\cite{PhysRevB.13.5188,10.1063/1.3382344}. Furthermore, hybrid functionals and the spin-orbit coupling (SOC) effect are taken into account to achieve high accuracy in electronic band gap calculations. For the calculation of the electronic band structure employing the plane wave basis set, preliminary geometric relaxation is performed with an energy cut-off value of 800 eV, a maximum atomic force of 0.005 eV/Å, and a pressure threshold of 0.05 GPa~\cite{PhysRevB.50.17953}. Rigorous dynamical matrix calculations were performed using an ultra basis set in a $5 \times 3 \times 1$ supercell to confirm the dynamical stability of the structure. 

To explore the ballistic phonon transport properties, the phonon transmission spectrum, $\tau_{ph}(\omega)$, is calculated using a $1\times 25 \times 1$ q-point sampling with a high energy resolution of 0.1 meV. Phonon transmission calculations are conducted based on the non-equilibrium Green's function method as follows~\cite{10.1063/1.3531573},
\begin{equation}
    \tau_{ph}(\omega) = Tr[\Gamma_L G_c^R \Gamma_R G_c^A]
\end{equation}
Here, $G_c^A=G{_c^R}^{\dagger}$ denotes the advanced Green's function, and $\Gamma_{L/R}$ are broadening matrices of the left and right electrodes.
Following the Landauer formalism, phonon thermal conductance $\kappa_{ph}$ is obtained by~\cite{PhysRevLett.81.232,Sevincli_2019},
\begin{eqnarray}
\kappa_{ph} = \frac{1}{2\pi} \int \tau_{ph}(\omega) \hbar \omega \left(\frac{\partial f_{BE} (\omega, T)}{\partial T}\right)d\omega
\label{kappaph_eq}
\end{eqnarray}
Electron transmission $T(E)$ can be derived through counting available transport channels under zero-bias voltage as implemented in QATK. This approach accurately captures the intrinsic, scattering-free ballistic limit by computing the exact number of available modes along the transport direction. 
Once the $T(E)$ is obtained, electronic thermoelectric coefficients i.e. electrical conductance $G_e=e^2 L_0$, Seebeck coefficient (thermopower) $S=(L_1/L_0)/eT$, power factor $PF = S^2G$ and electrons contribution to the total thermal conductance $\kappa_{el}=(L_2-L_1^{2}/L_0)/T$ can be evaluated through transport integrals, $L_n$ as following within the Landauer formalism~\cite{PhysRevB.33.551,PhysRevB.73.085406,Datta_1995},
\begin{eqnarray}
L_n(\mu,T) = -\frac{2}{h}\int T(E) (E-\mu)^{n} \left(-\frac{\partial f_{FD}}{\partial E}\right) dE
\end{eqnarray}

Unlike the zero-bias $T(E)$ used to reveal the thermoelectric coefficients, the current-voltage $I-V$ curves are derived by computing the $T(E)$ at finite bias voltages by utilizing the NEGF method~\cite{PhysRevB.65.165401}. So as to, a two-probe model consisting of left electrode, right electrode, and central region, is designed for transport calculations for armchair and zigzag directions. The electrical current is determined by computing the bias-dependent transmission function, $\mathcal{T}(E, V_{b})$. One can obtain $\mathcal{T}(E, V_{b})$ from $G(E,V_b)$ and so-called broadening functions $\Gamma^{L/R}$
\begin{equation}
\mathcal{T}(E,V_b)=Tr[G(E,V_b)\Gamma^LG^{\dagger}(E,V_b)\Gamma^R]    
\end{equation}
The electrical current is evaluated via integrating the $\mathcal{T}(E, V_{b})$ over the applied bias according to the Landauer-Büttiker formalism~\cite{PhysRevB.31.6207},
\begin{equation}
I(V_b)=\frac{2e}{h}\int_{-\infty}^{+\infty} \mathcal{T}(E_,V_b)[f_L(E,V_b)-f_R(E,V_b)dE]
\end{equation}
where $f_{L/R}(E,V_b)$ represent the Fermi-Dirac distribution functions of the left and right electrodes, respectively.
In transport simulations, k-point density is set to 1 $\times$ 4 $\times$ 150. To ensure convergence of the self-consistent cycles, the bias voltage was incremented in steps of 0.02 V and 0.025 V in the armchair and zigzag directions, respectively.
\begin{figure}[htbp]
\includegraphics[width=\linewidth]{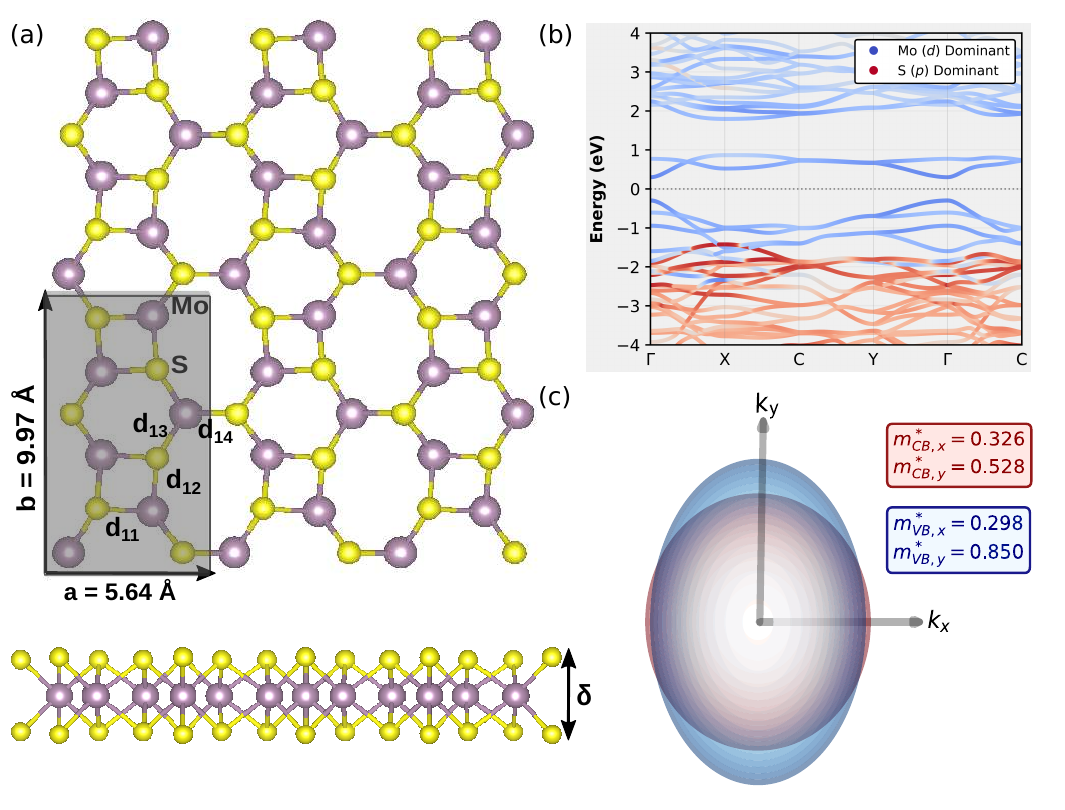}
\caption{\label{fig:1}(a) Atomic configuration of MoS$_{2}$-BPN from top and side view, (b) projected electronic band structure (fat band structure) and (c) schematic representation of anisotropic nature of carrier effective masses along armchair ($x$) and zigzag ($y$) directions.  }
\end{figure}

\begin{table*}[t]
\small
\centering
\caption{Optimized lattice constants $a$ and $b$, bond lengths $d_{ij}$, band gap values of MoS$_2$-BPN. The electronic band gaps are computed employing both the LCAO basis set (incorporating PBE, PBE+SOC, and HSE+SOC functionals) and the PW basis set.}
\label{Table:1}
\begin{tabular*}{\textwidth}{@{\extracolsep{\fill}}cccccccccc@{}} 
\toprule
\textbf{a/b} & \textbf{d$_{11}$} & \textbf{d$_{12}$} & \textbf{d$_{13}$} & \textbf{d$_{14}$} & \textbf{$\delta$} & $E_{g, \text{LCAO}}^{\text{PBE}}$ & $E_{g, \text{LCAO}}^{\text{PBE+SOC}}$ & $E_{g, \text{LCAO}}^{\text{HSE+SOC}}$ & $E_{g, \text{PW}}^{\text{PBE}}$ \\
(\AA) & (\AA) & (\AA) & (\AA) & (\AA) & (\AA) & (eV) & (eV) & (eV) & (eV) \\
\midrule
5.7/10.0 & 2.43 & 2.45 & 2.40 & 2.41 & 3.23 & 0.35 & 0.33 & 0.60 & 0.34 \\
\bottomrule
\end{tabular*}
\end{table*}

\section{Results and discussion}
\subsection{Structural and Electronic and Transport Properties}
\begin{figure}[htbp]
\includegraphics[width=\linewidth]{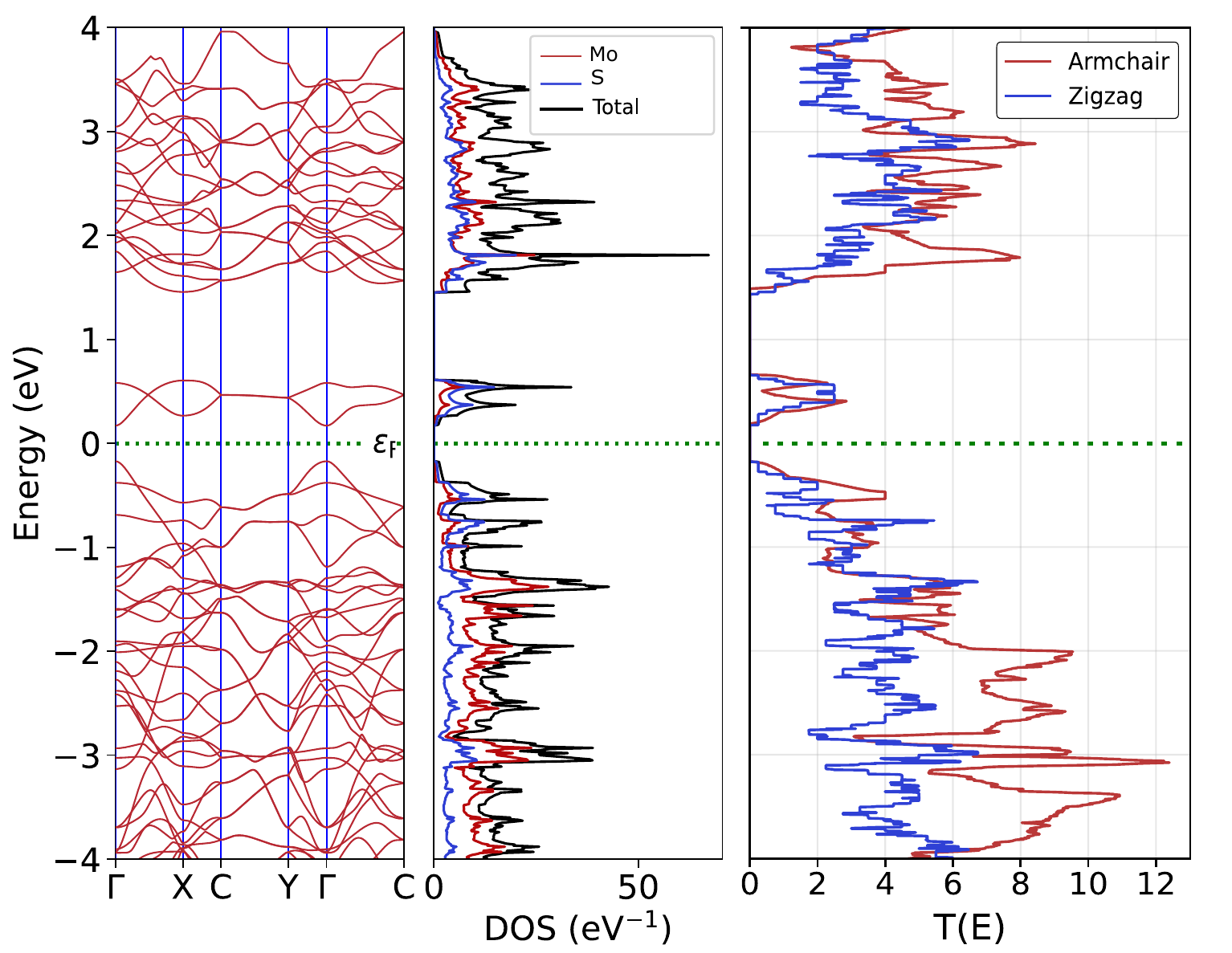}
\caption{\label{fig:elband_pdos_trans}Electronic band structure, element-based projected density of states and electronic transmission spectrum of MoS$_{2}$-BPN in armchair and zigzag directions.}
\end{figure}

As schematically shown in Figure~\ref{fig:1}, the fully relaxed geometry of MoS$_2$-BPN includes 18 atoms in total, incorporating 6 Mo and 12 S atoms due to the ($1\times2$) reconstruction along the $y$-axis. The zigzag ($y$)-direction is formed by parallel rows of mixed square-hexagon chains adjacent to pure octagon chains. In contrast, the armchair ($x$)-direction comprises connected square-octagon chains aligned next to pure hexagon rows. These polygonal rings undergo slight distortions compared to the planar C-BPN.
To provide the structural details of optimized geometry, the resulting lattice parameters and bond lengths are tabulated in Table~\ref{Table:1}. Unlike the C-BPN and its derivatives, which form an ultra-flat layer, the MoS$_2$-BPN possesses a certain thickness of $\delta=3.23~\text{\AA}$ due to its intrinsic S-Mo-S sandwiched crystal structure.
Electronic band structure and element-based projected density of states (PDOS) depicted in Figure~\ref{fig:elband_pdos_trans} are essential to evaluate the transport and thermoelectric properties of MoS$_2$-BPN. Within the GGA, MoS$_2$-BPN possesses a direct band character with a value of 0.35 eV by employing the PseudoDojo Ultra basis set. Spin-orbit coupling effect on the band gap value of MoS$_2$-BPN is significantly lower compared to that of the hexagonal MoS$_2$ counterpart~\cite{PhysRevB.84.153402}. The inclusion of SOC results in a negligible reduction of approximately 15 meV in the band gap value of MoS$_2$-BPN. The calculated band gap difference between the HSE+vdW+SOC and HSE+vdW functionals is found to be 0.76 meV, which is close to that computed by Demirci \textit{et al.}~\cite{doi:10.1021/acs.jpcc.3c00388}. For comparison, an additional band structure calculation using a plane-wave basis set was performed, yielding a band gap of 0.34 eV, which is consistent with previously reported values~\cite{doi:10.1021/acs.jpcc.3c00388}. To provide a detailed comparison, all calculated band gap values derived from various exchange functionals and basis sets are summarized in Table~\ref{Table:1}.

The lowest two conduction bands and topmost valence bands are dominated by $d$ orbitals of $Mo$ atoms, whereas the deeper conduction bands are governed by the $p$ orbitals of $S$ atoms, as evidenced by the fat band and DOS characters in Figure~\ref{fig:1} and \ref{fig:elband_pdos_trans}. Fundamentally, the transport properties of charge carriers in nano devices are partially dictated by the effective mass of the charge carriers at the VBM and CBM. Accordingly, electron ($m_e^*$) and hole effective masses ($m_h^*$) along armchair and zigzag directions are determined by $$m^* = \hbar^2 \left( \frac{\partial^2 E(k)}{\partial k^2}\Bigg|_{k=k_0} \right)^{-1}$$ based on a finite difference representation (using a step size of $\Delta k = 0.001$ Å$^{-1}$). In the armchair direction, calculated $m_e^*$ and $m_h^*$ are exceptionally smaller than those in the zigzag direction, as summarized in Table~\ref{Table:2}. 
In addition, electron ($\mu_e$) and hole mobilities ($\mu_h$) are further calculated for both directions, accounting for full electron-phonon coupling matrix elements at the band edges. In accordance with the effective masses, $\mu_e$ and $\mu_h$ in the armchair direction are drastically larger than those along the zigzag direction. Carrier mobilities exhibit a strong anisotropy; electrons and holes possess $2.7\times10^5$ and $1.4\times10^2$ $cm^2 s^{-1}V^{-1}$ in armchair geometry, while they drop to $9.4\times10^3$ and $9.2$ $cm^2 s^{-1}V^{-1}$ along zigzag direction, respectively. Specifically, while electrons in the armchair direction demonstrate remarkably high mobility that perfectly aligns with ballistic transport, holes in the zigzag direction are restricted to a sub-nanometer mean free path due to strong electron-phonon scattering. Therefore, it is worth noting that the ballistic model will yield the upper theoretical limit for the $p-type$ thermoelectric performance along the zigzag direction. 

\begin{table}[htbp]
  \centering
  \small
  \caption{The calculated values of effective masses and room temperature mobilities for electrons and holes along armchair and zigzag directions.}
  \label{Table:2}
  \begin{tabular}{l c c c c}
    \hline
    \textbf{Direction} & \textbf{m$_e^*$ ($m_0$)} & \textbf{m$_h^*$ ($m_0$)} & \textbf{$\mu_{e}$ (cm$^2$ s$^{-1}$ V$^{-1}$)} & \textbf{$\mu_{h}$ (cm$^2$ s$^{-1}$ V$^{-1}$)} \\
    \hline
    Armchair & 0.33 & 0.30 & 2.7$\times10^5$ & 1.4$\times10^2$ \\
    Zigzag   & 0.53 & 0.85 & 9.4$\times10^3$ & 9.2 \\
    \hline
  \end{tabular}
\end{table}

\begin{figure}[htbp]
\includegraphics[width=\linewidth]{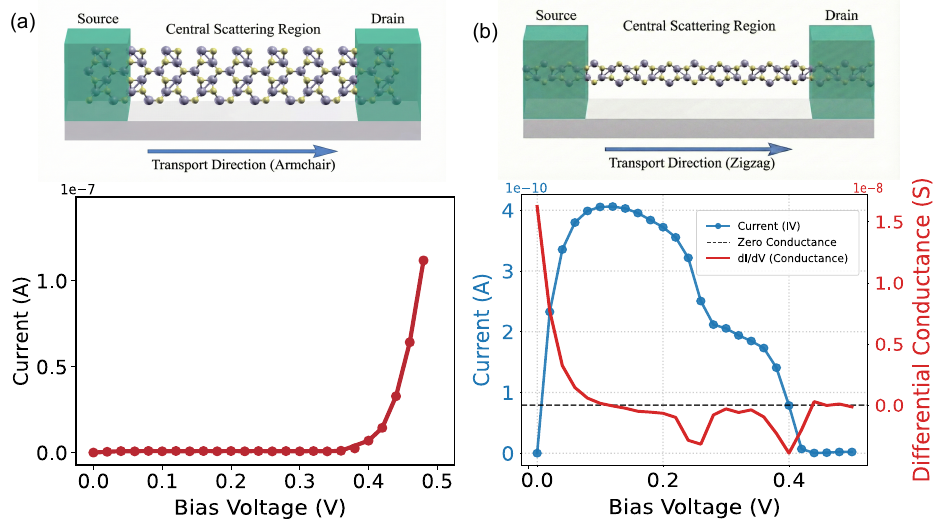}
\caption{\label{fig:IV_curve}Two-probe device models and current vs bias voltage trends along (a) armchair and (b) zigzag direction.}
\end{figure}

To explore the nano-device potential and carrier dynamics of MoS$_2$-BPN, current-voltage ($I-V$) characteristics are calculated for both directions utilizing NEGF+DFT method under finite bias voltages as plotted in Figure~\ref{fig:IV_curve}. The resulting $I-V$ curves reveal profound anisotropy in current magnitude as well as conduction mechanism are observed between armchair and zigzag directions. Current values reach $10^{-7}$A along the armchair direction, exhibiting conduction approximately three orders of magnitude ($10^3$ times) higher than in the zigzag direction. The high current-carrying capacity arises from high $\mu_e$ and $\mu_h$ values in the armchair direction. According to Figure~\ref{fig:IV_curve}(a), the calculated current along the armchair direction is suppressed while the bias window resides within the band gap, followed by a sharp increase when it surpasses the band gap value. In addition, the armchair configuration provides an intrinsic switching ON/OFF ratio of $\sim$317. However, the performance profile becomes even more pronounced in terms of spatial anisotropy, yielding a directional current contrast ($I_{armchair}/I_{zigzag}$) of approximately $7 \times 10^4$ at an applied bias of 0.50 V. This spatial ratio is highly comparable to the $10^4$–$10^5$ benchmarks recently reported for other highly anisotropic 2D materials, such as~\cite{doi:10.1021/acs.jpclett.1c03477,nano15090679}.
In contrast, distinct electron transport behavior is observed for zigzag orientation. As shown in Figure~\ref{fig:IV_curve}(b), current levels reduce to $10^{-10}$A in a zigzag direction, which falls drastically short of the armchair direction's performance. This behavior perfectly coincides with the carrier mobilities in the zigzag direction. Furthermore, the zigzag current rises at low biases, reaching its maximum at a bias of 0.12 V. Remarkably, current then decreases when the bias voltage takes values beyond 0.12 V, which indicates that the system enters a negative differential conductance (NDG) regime. The current falls sharply to a valley of $3.19 \times 10^{-13}$ A at 0.44 V, resulting in an exceptional Peak-to-Valley Ratio (PVR) of approximately 1270. These findings reveal that crystal orientation determines the overall behavior of the device, offering both high directional isolation and NDG within the same material.

\subsection{Phonon Transport Properties}
\begin{figure}[htbp]
\includegraphics[width=\linewidth]{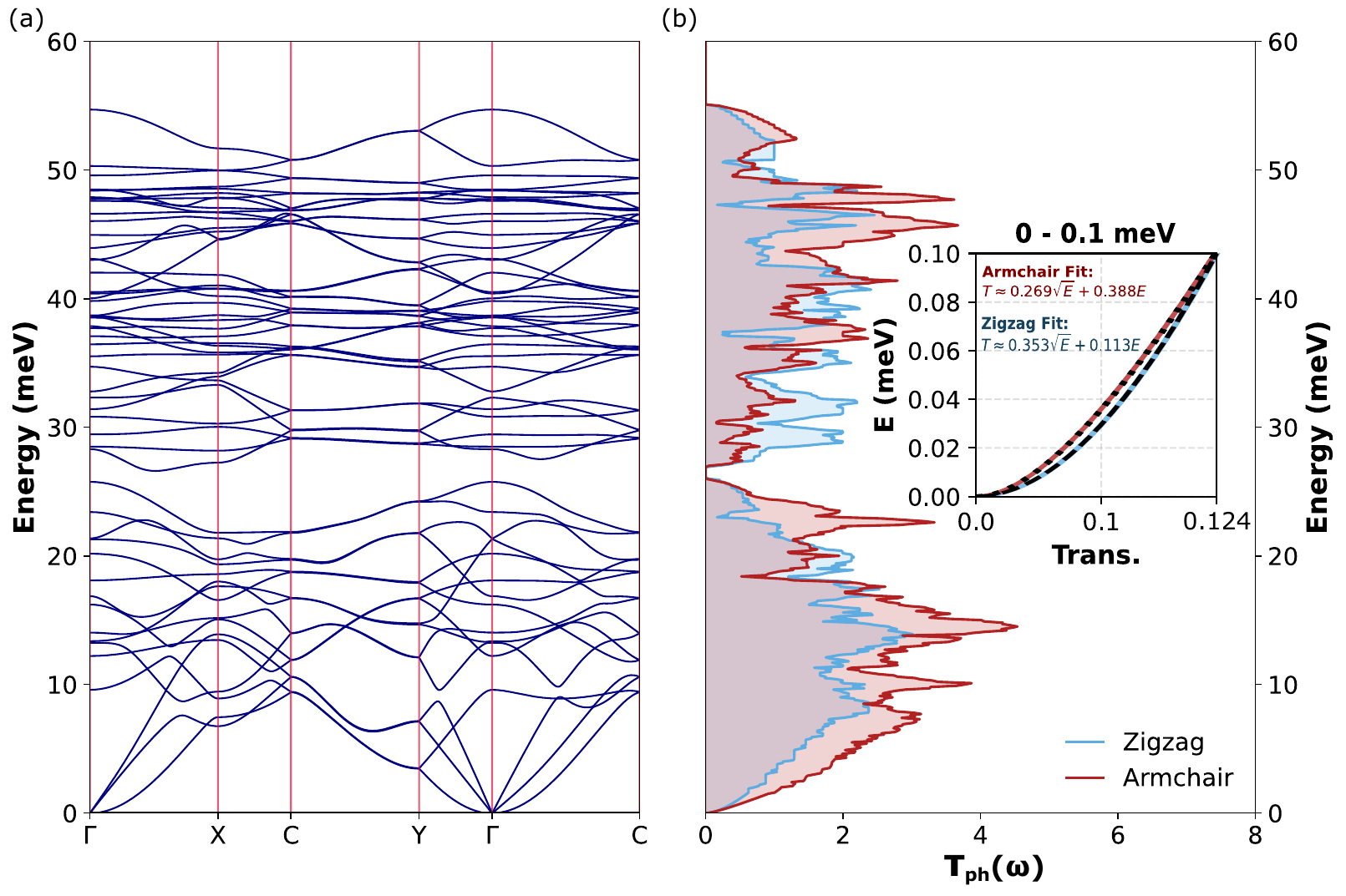}
\caption{\label{fig:ph_band_trans}(a) Phonon band structure and (b) phonon transmission spectrum of MoS$_{2}$-BPN along armchair and zigzag direction. Inset: Phonon transmission behavior in an extremely low energy range between 0 and 0.1 meV.}
\end{figure}

Absence of the imaginary frequencies in the phonon spectrum ensures the dynamical stability of MoS2-BPN at $T=0~\text{K}$ as shown in Figure~\ref{fig:ph_band_trans} (a). Since the MoS$_2$-BPN structure has 18 atoms in its unit cell, there are 54 phonon branches, of which three are acoustic, and 51 of them are optical modes. Whereas the maximum phonon energy of the hexagonal 2H-MoS$_2$ monolayer is approximately 58.6 meV~\cite{PhysRevB.84.155413}, this energy downshifts to 54.70 meV in the MoS$_2$-BPN structure. In the ballistic limit, the maximum phonon energy directly determines the phonon transmission energy range. A narrow phonon band gap arises between optical phonon bands, which is approximately 1 meV. The acoustic phonon branches provide critical insights into the mechanical anisotropy of the MoS$_2$-BPN monolayer. The longitudinal acoustic (LA) sound velocities reveal a pronounced in-plane anisotropy, with a significantly higher velocity along the zigzag direction ($v_{\mathrm{LA,y}} \approx 6.2\times 10^3$ m/s) than along the armchair direction ($v_{\mathrm{LA,x}} \approx 4\times 10^3 $ m/s). This indicates the zigzag direction of the MoS$_2$-BPN structure behaves more stiffly compared to the armchair direction. Furthermore, this is highly consistent with the elastic constants previously reported for MoS$_2$-BPN, confirming that the interconnected octagonal rings contribute dominantly to the in-plane mechanical robustness~\cite{doi:10.1021/acs.jpcc.3c00388}. Furthermore, the out-of-plane acoustic (ZA) branch allows for an assessment of the bending rigidity. In the long-wavelength limit near the $\Gamma$ point, the dispersion of the ZA mode follows a quadratic relation,$\omega_{ZA}=\alpha q^2$, where $\alpha$ is the curvature. Our calculations demonstrate that the ZA mode exhibits lower curvature along the zigzag direction ($\alpha_y \approx 3.53 \times 10^{-7}$ m$^2$/s) than the armchair direction ($\alpha_x \approx 5.36 \times 10^{-6}$ m$^2$/s). From these phonon characteristics, it can be inferred that the MoS$_2$-BPN possesses bending anisotropy. 

After revealing the phonon characteristics of MoS$_2$-BPN, the phonon transmission spectrum can be directly evaluated as presented in Figure~\ref{fig:ph_band_trans} (b). At extremely low energies, $\tau_{ph}(\omega)$ along the zigzag direction is noticeably higher than that along the armchair direction (Figure~\ref{fig:ph_band_trans} (b), inset). This directional dependence of $\tau_{ph}(\omega)$ originates directly from the ZA modes. The $ZA$ phonons possess a quadratic dispersion relation ($\omega = \alpha q^2$), yielding a transmission behavior proportional to $\sqrt{E}$. Higher structural out-of-plane flexibility along the zigzag direction (reflected by its much smaller dispersion coefficient, $\alpha_z \approx 3.53 \times 10^{-7}$ m$^2$/s compared to $\alpha_a \approx 5.36 \times 10^{-6}$ m$^2$/s), leads to a high density of phonon states near the $\Gamma$ point. Our theoretical fits ($T \approx A\sqrt{E} + BE$) confirm this, revealing a dominant $ZA$ coefficient for the zigzag direction ($A_z \approx 0.353$) compared to the armchair direction ($A_a \approx 0.269$). However, this zigzag-dominated regime is confined to the extreme low-energy spectrum. As the phonon energy increases, the linear transmission contributions from the in-plane acoustic modes (TA and LA, corresponding to the $BE$ term) become increasingly effective. Since the armchair direction is considerably stiffer in-plane (evidenced by its larger linear coefficient, $B_a \approx 0.388$ vs. $B_z \approx 0.113$), its transmission grows much more rapidly. Consequently, at a crossover energy of approximately $\sim 0.094$ meV, the strong in-plane (TA/LA) contributions of the armchair direction fully compensate for its initial ZA deficiency, causing the armchair transmission to overtake the zigzag direction and dominate the subsequent low-energy transport. Although transmission in the zigzag direction surpasses that in the armchair direction at mid-frequencies, when considering the entire phonon spectrum, $\tau_{ph}(\omega)$ in the armchair direction generally exhibits a higher trend than that in the zigzag direction. 

\begin{figure}[htbp]
\includegraphics[width=\linewidth]{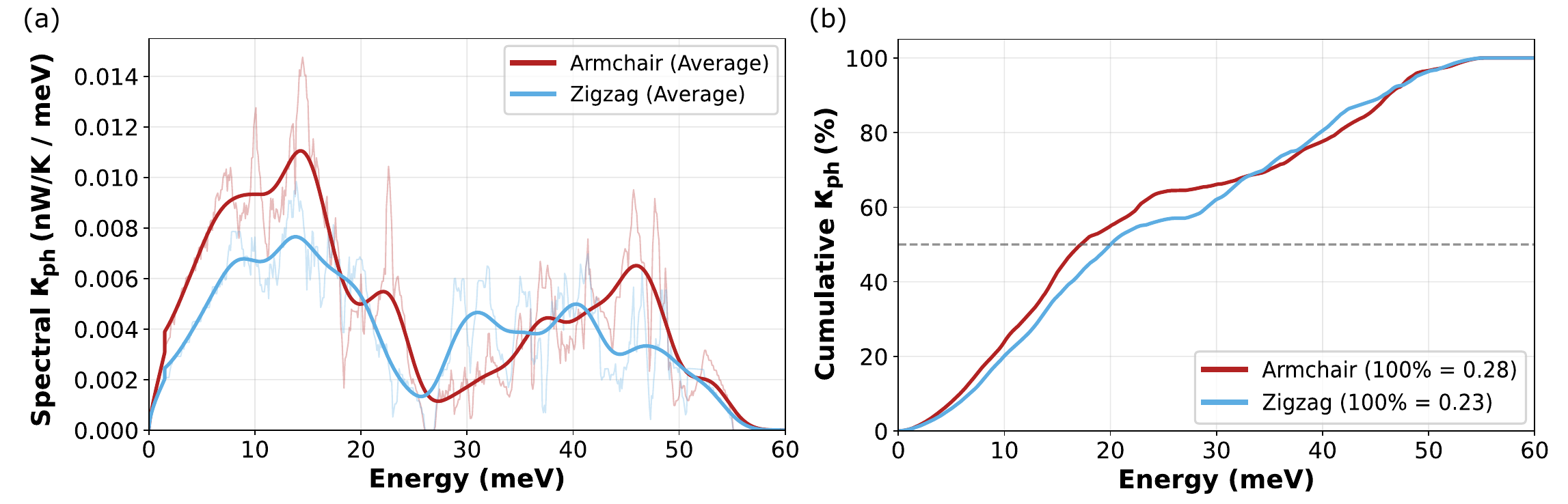}
\caption{\label{fig:ph_thermal_cond_spec}(a) Spectral and (b) cumulative phonon thermal conductance of MoS$_{2}$-BPN in the armchair and zigzag orientation at 300~K.}
\end{figure}

Phonon thermal conductance, which directly affects the TE efficiency of MoS$_2$-BPN is obtained based on Eq.~\ref{kappaph_eq}. Before discussing the total $\kappa_{ph}$ profile, spectral phonon conductance can provide insight on anisotropic behavior in heat transport as presented in Figure~\ref{fig:ph_thermal_cond_spec} (a). For both directions, the primary contribution to phonon thermal transport comes from low-energy acoustic phonons in the 0–20 meV range. Similar to transmission behavior, spectral conductance along the armchair direction exceeds that along the zigzag direction until mid-energy ranges. Spectral conductance in the zigzag direction surpasses that in the armchair direction between 28 meV and 38 meV as in the transmission spectrum. The cumulative thermal conductance (Figure~\ref{fig:ph_thermal_cond_spec} (b)) shows that 50$\%$ of the total heat is carried by phonons with energies below $\sim$18 meV. This indicates that enhancing the TE efficiency of MoS$_2$-BPN can be achieved through nanoengineering strategies that specifically target low-energy phonons.

\begin{figure}[htbp]
\includegraphics[width=\linewidth]{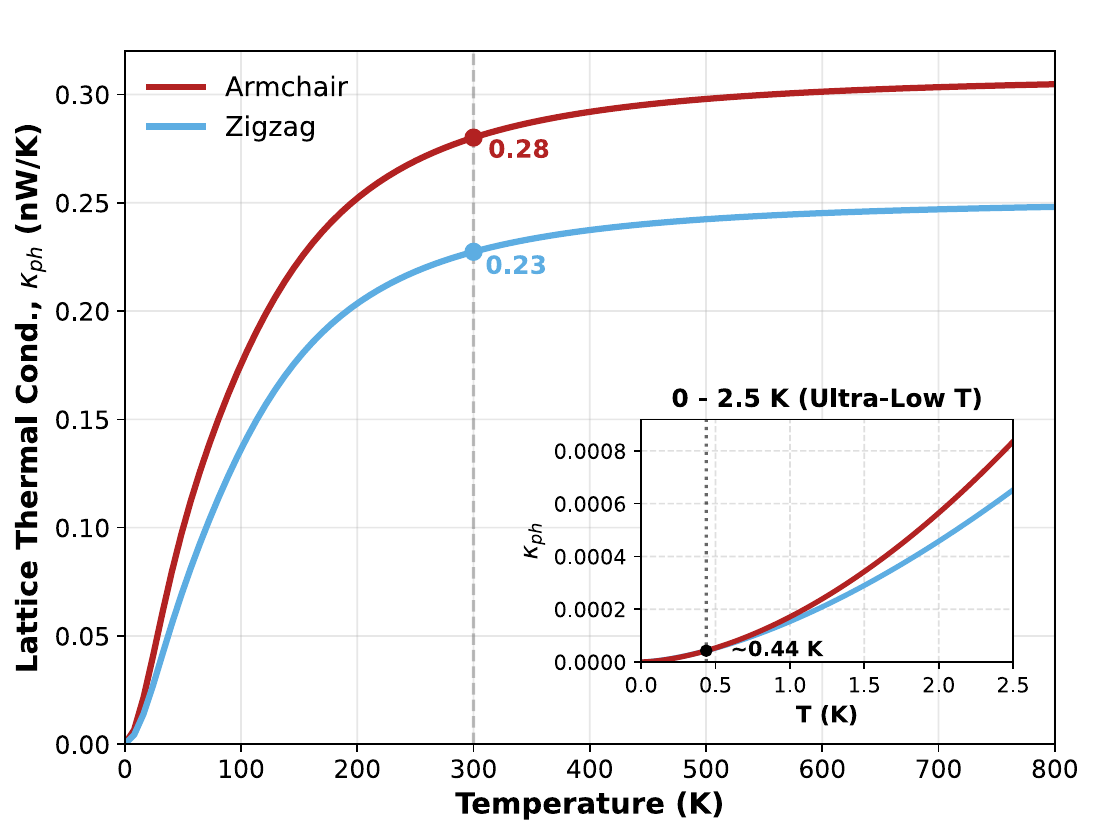}
\caption{\label{fig:ph_thermal_cond}Variation of the phonon thermal conductance of MoS$_{2}$ BPN as a function of temperature along armchair and zigzag orientation. Inset: Profile of $\kappa_{ph}$ in the ultra-low temperature range for both directions. }
\end{figure}

Figure~\ref{fig:ph_thermal_cond} reveals that the total $\kappa_{ph}$ rises rapidly with the increase in temperature due to the excitation of phonon modes. However, all phonon modes up to the maximum frequency are fully populated at elevated temperatures, $\kappa_{ph}$ reaches saturation as expected in the ballistic phonon transport limit. Phonon thermal conductance of an armchair direction (0.28 nW/K) outperforms that of the zigzag direction (0.23 nW/K) as demonstrated in Figure~\ref{fig:ph_thermal_cond}. Since the Bose-Einstein thermal weighting function favors low-energy acoustic phonon modes and filters out high-energy modes, the directional anisotropy in $\kappa_{ph}$ appears weaker than in phonon transmission. It should be noted that while the crossover in the $\tau_{ph}(\omega)$ occurs around $0.1$ meV (Figure~\ref{fig:ph_band_trans} (b), inset), the corresponding crossover in $\kappa_{ph}$ appears at $0.44$ K ($k_B T \approx 0.038$ meV) as shown in inset of the Figure~\ref{fig:ph_thermal_cond}. This apparent energy shift is a direct consequence of the Bose-Einstein thermal weighting function, $E^2 \partial f_{BE}/\partial T$, which peaks at energies significantly higher than $k_B T$ ($\sim 2.5 - 3.0 k_B T$), directly mapping the $0.44$ K thermal profile to the $0.1$ meV transmission features.

\begin{table*}[t]
\small
\caption{Maximum thermoelectric figure of merit ($ZT_{\text{max}}$) near the fermi level and corresponding optimal TE transport ($S$), electronic conductance ($G_e$), electronic thermal conductance ($\kappa_{\text{el}}$) and total thermal conducance ($\kappa$) at 300 K for the armchair and zigzag orientations.}
\label{tab:te_params}
\begin{tabular*}{\textwidth}{@{\extracolsep{\fill}}lccccccc}
\toprule
Orientation & Doping & $G_e$ ($G_0$) & $S$ (mV/K) & $PF$ ($10^{-3}$ nW/K$^2$) & $\kappa_{\text{el}}$ (nW/K) & $\kappa$ (nW/K) & $ZT_{\text{max}}$ \\
\midrule
Armchair & $p$-type & 0.18 & 0.20  & 0.28 & 0.03 & 0.31 & 0.27 \\
Armchair & $n$-type & 0.18 & $-0.20$ & 0.27 & 0.04 & 0.32 & 0.26 \\
Zigzag   & $p$-type & 0.64 & 0.10  & 0.25 & 0.17 & 0.39 & 0.19 \\
Zigzag   & $n$-type & 0.79 & $-0.10$ & 0.30 & 0.21 & 0.44 & 0.20 \\
\bottomrule
\end{tabular*}
\end{table*}

\subsection{Thermoelectric Properties}
\begin{figure}[htbp]
\includegraphics[width=\linewidth]{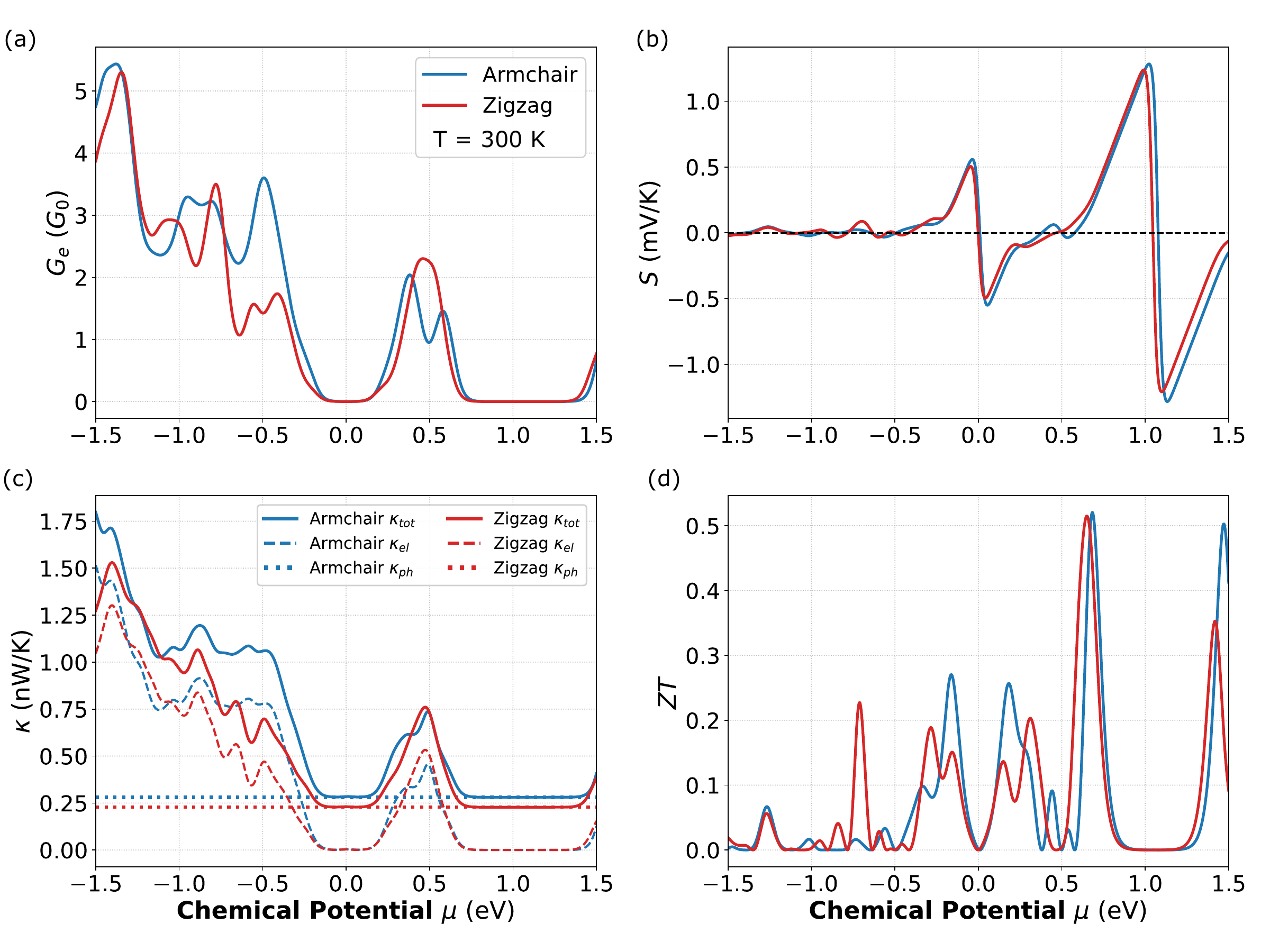}
\caption{\label{fig:TE_coeff}Variation of the thermoelectric coefficients of MoS$_{2}$-BPN as a function of chemical potential along the armchair and zigzag directions at room temperature: (a) electronic conductance, (b) Seebeck coefficient, (c) electronic, phonon, and total thermal conductances, and (d) thermoelectric figure of merit ($zT$).}
\end{figure}

Once the electron transmission spectrum is obtained based on NEGF formalism, the electronic TE coefficients are evaluated using the Landauer transport integrals defined in Section~\ref{Method_sec}. As illustrated in Figure~\ref{fig:TE_coeff}(a), electronic conductance can be interpreted as a thermally broadened electronic transmission function. As summarized in Table~\ref{tab:te_params}, specifically at the optimum chemical potential that maximizes $zT$, the $p-type$ $G_e$ in the zigzag direction significantly exceeds that of the armchair direction, highlighting a strong transport anisotropy. 
At the optimum chemical potential, the $p$-type and $n$-type $G_e$ exhibit similar values for the armchair direction, while at higher chemical potentials, the $p$-type $G_e$ reaches a significantly higher value than the $n$-type $G_e$. Regarding the Seebeck coefficient, armchair direction has a higher $p$- and $n-type$ $S$ of 0.2 mV/K. In addition, peak values of $p$- and $n-type$ $S$ reach 0.56/0.50 mV/K within the band gap region and exhibit a symmetric behavior around the Fermi level as depicted in Figure~\ref{fig:TE_coeff}(b). Combined effect of the $G_e$ and $S$ yields highly competitive $PF$ values for both transport directions. Therefore, the decisive determining factor for the TE performance is the $\kappa_{el}+\kappa_{ph}=\kappa$ plotted in Figure~\ref{fig:TE_coeff}(c). $\kappa_{el}$ follows an analogous trend with the $G_e$ as expected from the Wiedemann-Franz law, $\kappa_{el} \propto G_e T$. As seen in Table~\ref{tab:te_params}, the total thermal conductance, $\kappa$, exhibits lower values along the armchair direction, making it highly favorable for thermoelectric applications. Overall TE efficiency, $p/n-type$ $zT$ values reach 0.27/0.26 and 0.19/0.20 in the vicinity of the Fermi level along armchair and zigzag directions, respectively, which signifies the armchair direction shows preferable TE performance compared to the zigzag direction. 

\section{Conclusions}
This study systematically explores the electronic, thermal, thermoelectric, and transport properties of MoS$_2$-biphenylene using density functional theory combined with NEGF formalism. Ballistic thermal transport calculations reveal that phonon thermal conductance at 300~K is notably suppressed in both directions, yielding values of 0.28 nW/K for the armchair and 0.23 nW/K for the zigzag configurations. The concurrent anisotropy in both electronic and phononic transport leads to distinct room-temperature p- and n-type $zT$ values of 0.27/0.26 and 0.19/0.20 for the armchair and zigzag directions, respectively. Beyond revealing the TE properties of MoS$_2$-BPN, quantum transport simulations are also conducted to elucidate the current-voltage characteristics. The exceptionally high current magnitudes along the armchair direction observed in this regime highlight the material's strong potential for high-performance nanoscale device applications. In terms of directional anisotropy, $I_{armchair}/I_{zigzag} = 7 \times 10^4$ offers a robust switching mechanism for nano-device applications. Conversely, transport along the zigzag direction is characterized by a pronounced intrinsic negative differential conductance (NDC). In this aspect, MoS$_2$-biphenylene stands out as a highly promising and manufacturing-friendly candidate for next-generation logic and oscillator devices.

\section*{Author Contributions}
G.Ö.S. is the sole author of this work and assumes full responsibility for the conceptualization of the study, execution of the computational simulations, formal data analysis, and the writing and editing of the manuscript.

\section*{Conflicts of interest}
There are no conflicts to declare.

\section*{Data availability}
The datasets generated and analysed during the current study are available from the corresponding author upon reasonable request.

\section*{Acknowledgements}
G.Ö.S. acknowledges support from the TUBITAK project 123C159. Part of the numerical calculations is carried out at TUBITAK ULAKBIM, High Performance and Grid Computing Center (TRUBA resources).

\end{document}